\documentclass[aps,pra,longbibliography,amsmath,amssymb,notitlepage]{revtex4-1}

\usepackage{hyperref}
\usepackage{color}
\usepackage{rotating}

\begin{document}

\title{Inertial migration of a deformable particle in pipe flow}
\author{Dhiya Alghalibi$^{1,2}$}
\author{Marco E. Rosti$^1$}
\email[Corresponding author: ]{merosti@mech.kth.se}
\author{Luca Brandt$^1$}
\affiliation{$^1$Linn$\acute{\textrm{e}}$ Flow Centre and SeRC (Swedish e-Science Research Centre), \\ KTH Mechanics, S-100 44 Stockholm, Sweden \\
$^2$College of Engineering, University of Kufa, Al Najaf, Iraq}

\begin{abstract}
We perform fully Eulerian numerical simulations of an initially spherical hyperelastic particle suspended in a Newtonian pressure-driven flow in a cylindrical straight pipe. We study the full particle migration and deformation for different Reynolds numbers and for various levels of particle elasticity, to disentangle the interplay of inertia and elasticity on the particle focusing. We observe that the particle deforms and undergoes a lateral displacement while traveling downstream through the pipe, finally focusing at the pipe centerline. We note that the migration dynamics and the final equilibrium position are almost independent of the Reynolds number, while they strongly depend on the particle elasticity; in particular, the migration is faster as the elasticity increases (i.e. the particle is more deformable), with the particle reaching the final equilibrium position at the centerline in shorter times. Our simulations show that the results are not affected by the particle initial conditions, position and velocity. Finally, we explain the particle migration by computing the total force acting on the particle and its different components, viscous and elastic.
\end{abstract}


\maketitle

\section{Introduction}

Particle (a general term to represents beads, capsules, vesicles, red blood cells,  etc.) migration in pipe and channel flows of different kind of fluids, Newtonian and non-Newtonian, has been extensively studied because of its importance in many industrial, environmental and biomedical applications. In these applications, the flow field and the dynamics of the particles motion are controlled by many parameters, such as the flow conditions, the carrier fluid, wall effects, inertial effects and particle deformability. The interplay between these various effects results in different interesting phenomena, such as particle separation \cite{pamme2007,karimi2013,lim2014} and focusing \cite{ateya2008,xuan2010,kang2013,li2015,lu2017}. These phenomena have been successfully applied for the manipulation of particles and cells in microfluidic devices. In this study we employ a fully Eulerian numerical algorithm based on the one-continuum formulation to fully resolve the fluid-structure interactions and the stresses in the liquid and solid phases and to provide an accurate understanding of the mutual effects of the inertia and particle elasticity on the motion of a deformable particle in a pipe flow, in conditions of interest for inertial microfluidic devices.

In a Newtonian fluid flow, the two most important non-dimensional parameters characterizing the deformable particle motion are the Reynolds and Weber numbers, quantifying inertia and particle elasticity, respectively. The Reynolds number $Re$ is defined as the ratio between inertial to viscous effects of the flow, while the ratio of inertia to elastic effects acting on the deformable particle is represented by the Weber number $We$. Generally, in the absence of inertia ($Re \approx 0$, i.e., Stokes flow), a neutrally buoyant rigid sphere follows the fluid motion without any lateral migration in order to satisfy the reversibility property of the Stokes flow \cite{guazzelli2011}. On the other hand, deformable particles in the same condition move towards low shear gradient regions, hence, when suspended in a Poiseuille flow they migrate towards the center of the channel \cite{kaoui2008}. The dynamics of deformable particles has mostly been investigated at low Reynolds numbers in the past (see, for example,\cite{risso2006,bagchi2007,coupier2008,lazaro2014,bacher2017}). Unlike Stokes flow, inertial flows are described by nonlinear governing equations, i.e., the flow system is irreversible. Thus, both rigid and deformable particles trajectories do not necessarily follow the behavior observed in the Stokes regime and particles undergo lateral movement. This is the case for typical inertial microfuidics applications ($Re > 1$ and $We > 0$), when inertial and elastic forces dominate the cross-streamline migration and final equilibrium position of the particles. In particular, elasto-inertial microfluidics is emerging as a powerful tool and research area, with devices where elasticity and inertia are being engineered to achieve efficient particle focusing and/or particle sorting \citep{di2018}.

Lateral migration and focusing of rigid particles were first observed experimentally in a Newtonian circular pipe flow by Segr{\`e} and Silberberg \cite{segre1961}. In a pipe flow, initially randomly distributed neutrally buoyant spheres immersed in a Newtonian carrier fluid migrate radially and focus into a narrow annulus at around $0.6$ the pipe radius, resulting in the so-called ``tubular pinch" effect. Later on, this effect has been carefully studied in several others analytical  \cite{schonberg1989,asmolov1999}, numerical \cite{feng1994,yang2005} and experimental  \cite{karnis1966,matas2004} investigations. Recently, an analogous effect was observed in laminar flows in rectangular and square-shape channels $(1 < Re < 2300 )$ \cite{di2007,di2009,choi2011}. Numerous studies which adapted this phenomenon to microfluidics applications, described it as ``inertial focusing" of particles. Indeed, the equilibrium position of the particles is the net result of two opposing forces resulting from the resistance of the solid particle to the deformation: \textit{(i)} the shear gradient lift force, which is induced by the velocity profile curve, that directs the particle away from the channel centerline towards the wall and \textit{(ii)} the wall-induced lift force arising from the interaction of the particle and the neighboring wall, that directs the particle away from the wall towards the channel centerline \cite{ho1974,schonberg1989,asmolov1999,martel2014}. These two competing forces, determining the lateral trajectory and the final equilibrium position of the particle, are modified differently by the blockage ratio \cite{di2009} and the flow Reynolds number \cite{matas2004}, and thus, by properly designing the geometry of microfluidics device, the lateral motion can be applied and cell focusing, separation, trapping, sorting, enrichment and filtration achieved (see the review articles by Di Carlo\cite{di2009} and Karimi et al.\cite{karimi2013}). Recent simulations from our group reported the mechanism of inertial focusing of both spherical and oblate particles in  microfluidics channels showing the entire migration dynamics of a particle from their initial to final equilibrium position, including particle trajectory, velocity, rotation and orientation \cite{lashgari2017}.

When the particle is deformable, the dynamics of the particle is further complicated by an additional force called ``deformation-induced lift force" arising from the deformation of the particle shape itself, that moves the particle towards the centerline and which becomes stronger as the particle deformation increases\cite{raffiee2017, hadikhani2018}. It is worth noting that, the alterations of the particle shape also affect and modify the two forces discussed previously, making the problem fully coupled. During the last 10 years, the dynamics of deformable particles has been studied both numerically \cite{doddi2008,salac2012,zhu_rorai_mitra_brandt_2014a,kim2015,wang2016} and experimentally \cite{mach2010,hur2011}. In particular, Hur et al.\cite{hur2011} showed that particles can be separated depending on their size and elastic deformability; same behaviour was also observed in numerical simulations \cite{kilimnik2011,chen2014}. In spite of the fact that all the results agree that the soft deformable particles move to the channel centerline, the effect of the flow Reynolds number is not quite understood and still debated. Indeed, while in some cases the final distance of the equilibrium position of the particle from the centerline appears to depend on the flow Reynolds number \cite{shin2011,kruger2014}, Kilimnik et al.\cite{kilimnik2011} found no proof of such a behaviour in their numerical simulations. However, they demonstrated that the distance of the final equilibrium position collapses on a single master curve when plotted versus the particles deformability \cite{leshansky2007,yang2011}. 

Most of the previous works on the dynamics of deformable particles in Newtonian flows in cylindrical straight pipes were mainly focused on low Reynolds number regimes. In addition, it is challenging to capture in experiments the entire migration dynamics of a deformable particle, such as its trajectories, the deformed shape and  the forces acting on the particle. Therefore, fully interface-resolved numerical simulations, where the full interactions between the solid and fluid phases are taken into consideration, can became a valuable tool to explore the problem. Also, numerical simulation may provide information about the role of various control parameters on the particle migration, such as the particle elasticity, the Reynolds number and the particle initial position.
 
In this work, we investigate the motion of an hyper-elastic deformable particle immersed in a Newtonian Poiseuille flow in a cylindrical straight pipe in different conditions. We focus on the lateral motion of the particle and compare the whole migration dynamics, the trajectory and the final equilibrium shape of the particle to shed more light onto the particle lateral displacement mechanism and to provide useful knowledge for the design of a microfluidic cylindrical system at finite inertia. In this manuscript, we investigate the effects of inertia and elasticity on the migration dynamics and equilibrium position of the particle. The paper is organised as follows. In $\S$2, we introduce  the governing equations, numerical method and simulations setup; the Results are presented in $\S$3, and finally, the main Conclusions and final remarks are summarised in $\S$4.

\begin{figure}[t]
  \centering
  \includegraphics[width=0.5\linewidth]{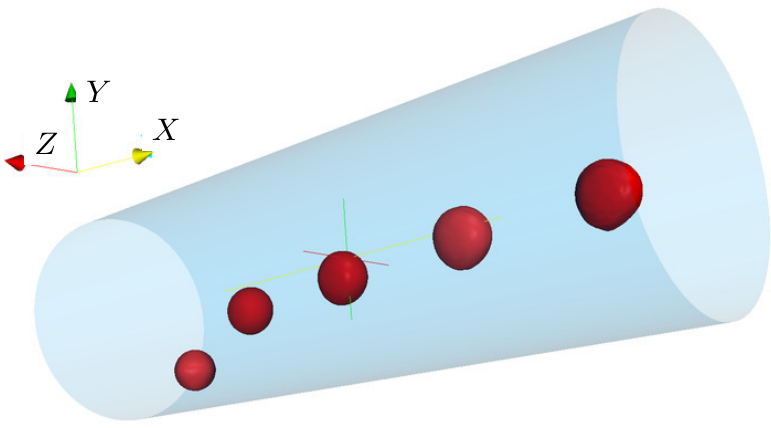}
  \caption{Visualization of the motion of a single deformable particle, initially spherical with radius $r$, in a cylindrical straight pipe of radius $R$, for the case $Re=200$ and $We=0.5$. The initial radial particle position is at $r=0.7487R$. The particle and the pipe are shown at the actual scale.}
  \label{fig:box} 
\end{figure}

\section{Methodology}
In this section we discuss the governing equations, the numerical method used to solve them and the setup to study the motion of a single deformable viscous hyperelastic particle suspended in a Newtonian pressure-driven Poiseuille flow in a straight cylindrical pipe geometry. 
 
\subsection{Governing equations}
In this work, we consider a deformable viscous hyperelastic particle immersed in a Newtonian viscous fluid. Hyper-elastic materials show non-linear stress-strain curves and are generally used to describe gel- and rubber-like substances; for example, \citet{verma2013} found a good agreement between experimental and numerical results where a soft gel is modelled as an incompressible viscous hyper-elastic material such as the one used here.

In the present work, both the fluid and solid phases are incompressible and their motion is governed by the momentum conservation equation and the incompressibility constraint
\begin{subequations}
\label{eq:NS}
\begin{align}
\rho \left( \frac{\partial u_i^{\rm f}}{\partial t} + \frac{\partial u_i^{\rm f} u_j^{\rm f}}{\partial x_j} \right) &=  \frac{\partial \sigma_{ij}^{\rm f}}{\partial x_j}, \\
\frac{\partial u_i^{\rm f}}{\partial x_i} &= 0, \\
\rho \left( \frac{\partial u_i^{\rm s}}{\partial t} + \frac{\partial u_i^{\rm s} u_j^{\rm s}}{\partial x_j} \right) &=  \frac{\partial \sigma_{ij}^{\rm s}}{\partial x_j}, \\
\frac{\partial u_i^{\rm s}}{\partial x_i} &= 0,
\end{align}
\end{subequations}
where $u$, $v$ and $w$ ($u_1$, $u_2$, and $u_3$) are the velocity streamwise and the two cross flow components, corresponding to the $x$, $y$ and $z$ ($x_1$, $x_2$, and $x_3$) coordinate directions, respectively (see figure~\ref{fig:box}). The superscripts $^{\rm f}$ and $^{\rm s}$ in the previous equations are used to distinguish the fluid and solid phases, and $\sigma_{ij}$ is the Cauchy stress tensor. Note that, the density of the fluid and solid phases $\rho$ is equal since we are considering a neutrally buoyant particle. The kinematic and dynamic interactions between the two phases are obtained by imposing the continuity of the velocity (i.e., the no-slip and no-penetration boundary conditions) and of the traction force (i.e., a traction balance) at the interface, i.e.,
\begin{subequations}
\label{eq:bc}
\begin{align}
u_i^{\rm f} &= u_i^{\rm s}, \label{bc-v}\\
\sigma_{ij}^{\rm f} n_j &=  \sigma_{ij}^{\rm s} n_j \label{bc-sigma},
\end{align}
\end{subequations}
where $n_i$ represents the normal vector at the interface. Finally, we need to define the Cauchy stress tensor for the fluid and solid systems; here, the carrier fluid is assumed to be Newtonian,
\begin{equation}
\label{eq:stress-f}
\sigma_{ij}^{\rm f} = -P \delta_{ij} + \mu^{\rm f} \left( \frac{\partial u_i^{\rm f}}{\partial x_j} + \frac{\partial u_j^{\rm f}}{\partial x_i} \right),
\end{equation}
while the particle is an incompressible viscous hyper-elastic material experiencing only the isochoric motion with constitutive equation
\begin{equation}
\label{eq:stress-s}
\sigma_{ij}^{\rm s} = -P \delta_{ij} + \mu^{\rm s} \left( \frac{\partial u_i^{\rm s}}{\partial x_j} + \frac{\partial u_j^{\rm s}}{\partial x_i} \right) + G \xi_{ij},
\end{equation}
where $\delta_{ij}$ is the Kronecker delta, the pressure is denoted by $P$ and the dynamic viscosity of the fluid and solid phases are indicated by $\mu^{\rm f}$ and $ \mu^{\rm s}$, respectively. The last term, $G \xi_{ij}$, is the hyperelastic stress contribution modeled as a neo-Hookean material, where $G$ indicates the modulus of transverse elasticity and $\xi_{ij}$ is the deviatoric left Cauchy-Green deformation tensor (also sometimes called Finger deformation tensor) defined as $\xi = F F^T$, where $F_{ij} = \partial x_i / \partial X_j$ is the deformation gradient (being $X$ and $x$ the initial and current coordinates \cite{bonet1997}). The previous set of equations for the solid material can be closed by updating the left Cauchy-Green deformation tensor components with the following transport equation:
\begin{equation}
\label{eq:B-adv}
\frac{\partial \xi_{ij}}{\partial t} + \frac{\partial u_k \xi_{ij}}{\partial x_k} - \xi_{kj}\frac{\partial u_i}{\partial x_k} - \xi_{ik}\frac{\partial u_j}{\partial x_k} = 0;
\end{equation}
this equation states that the upper convective derivative of $\xi$ is identically equally to zero, which is always true for an hyperelastic material \cite{bonet1997}.

\subsection{Numerical method}
To numerically solve the fluid-structure interaction problem at hand, we employ the so called one-continuum formulation \cite{tryggvason2007}, where only one set of equations is solved over the entire field. This is found by introducing a monolithic velocity vector field, $u_i$, valid everywhere; this is the weighted average between the values in the two phases, with the weight being a phase indicator function $\psi$ based on the local solid volumetric fraction in each cell \cite{quintard1994,takeuchi2010}
\begin{equation}
\label{eq:velocity}
u_i = \left( 1-\psi \right) u_i^{\rm f} + \psi u_i^{\rm s}.
\end{equation}
Thus, $\psi = 0$ or $\psi = 1$ if a computational cell in the domain is located inside the fluid or in the solid phase, while $\psi$ assumes a value between $0$ and $1$ in the interface cells. Thus, the governing equations (\ref{eq:NS}) can be rewritten as
\begin{subequations}
\label{eq:NSa}
\begin{align}
\rho \left( \frac{\partial u_i}{\partial t} + \frac{\partial u_i u_j}{\partial x_j} \right) &=  \frac{\partial \sigma_{ij}}{\partial x_j}, \\
\frac{\partial u_i}{\partial x_i} &= 0,
\end{align}
\end{subequations}
where the Cauchy stress tensor $\sigma_{ij}$ is written in a mixture form and defined as 
\begin{equation}
\label{eq:phi-stress}
\sigma_{ij} = \left( 1 - \psi \right) \sigma_{ij}^{\rm f} + \psi \sigma_{ij}^{\rm s}.
\end{equation} 
Finally, the local solid volume fraction $\psi$ is found by solving an additional transport equation
\begin{equation}
\label{eq:PHI-adv}
\frac{\partial \psi}{\partial t} + \frac{\partial u_k \psi}{\partial x_k} = 0.
\end{equation}

We solve equations (\ref{eq:NSa}), (\ref{eq:B-adv}) and (\ref{eq:PHI-adv}) in a fully Eulerian formulation on a staggered uniform mesh with velocities located on the cell faces and all the other variables (pressure, fluid and solid stress components) at the cell centers, as first proposed by \cite{sugiyama2011}. The time integration is based on an explicit fractional-step method \cite{kim1985}, where only the solid hyperelastic contribution in equation (\ref{eq:NS}) is advanced with the Crank-Nicolson scheme, while, all the other terms are advanced with the third order Runge-Kutta scheme \cite{min2001}. All the spatial derivatives are approximated with the second-order centred finite differences scheme, except for the advection term in equations (\ref{eq:B-adv}) and (\ref{eq:PHI-adv}) where the fifth-order weighted essentially non-oscillatory scheme is applied (\cite{shu2009, sugiyama2011, shahmardi2019}). A comprehensive review on the effect of different discretization schemes for the advection terms was studied by \cite{min2001}. The pressure is computed by solving the Poisson equation using fast Fourier transforms. In summary, the set of governing equations are solved as follows (see \cite{rosti2017}): (i) the left Cauchy-Green deformation tensor $\xi_{ij}$ and the local solid volume fraction $\psi$ are updated first by solving Equations \ref{eq:B-adv} and  \ref{eq:PHI-adv} (update step); (ii) the conservation of momentum equation \ref{eq:NSa} are advanced in time by first solving the momentum equation (prediction step), then by solving a Poisson equation for the projection variable and finally by correcting the pressure and velocity to ensure that the velocity field is divergence free (correction step).

The accuracy and validity of the code has been extensively examined in previous works, and more details on the numerical scheme and validation campaign are reported in Refs. \cite{rosti2017,rosti_brandt_2018a,rosti_brandt_mitra_2018a}, where very good agreement with literature results is obtained for various test cases. In addition, for more details on the numerical method, the reader is referred to Ref. \cite{sugiyama2011}.

\subsection{Simulations setup} 
\begin{table}
\centering
  \begin{tabular}{|c|c|c|c|c|c|c|}\hline
      $Re$   & $100$ & 200   & 400    &$r/R $   & $\mu^s/\mu^f$ \\[3pt]\hline

     $We$    & 0.125 & 0.125 & 0.125  & 0.2828  & 1.0   \\
             & $-$   & $-$   & 0.250  & 0.2828  & 1.0     \\
             & 0.500 & 0.500 & 0.500  & 0.2828  & 1.0     \\
             & $-$   & $-$   & 1.000  & 0.2828  & 1.0     \\
             & $-$   & $-$   & 2.000  & 0.2828  & 1.0     \\
             & 4.000 & 4.000 & 4.000  & 0.2828  & 1.0     \\
             & $-$   & $-$   & 1.000  & 0.7487  & 1.0     \\
             & $-$   & $-$   & 1.000  & 0.0544  & 1.0    \\
             & $-$   & $-$   & 4.000  & 0.2828  & 0.3     \\
             & $-$   & $-$   & 4.000  & 0.2828  & 3.0     \\\hline

  \end{tabular}
  \caption{Summary of all the simulations performed in the present study.}
  \label{tab:allcases}
\end{table}
In this study the pressure-driven motion of a hyperelastic deformable particle is examined in a circular straight pipe. The numerical domain is a square duct where the periodic boundary condition is applied in the streamwise $x$ direction, as shown in figure~\ref{fig:box}; the square duct is converted into a pipe via a volume penalization technique \cite{kadoch2012} that enforces the no-slip and no-penetration boundary conditions on the inner surface of the pipe. The pipe of diameter $D$ is $6D$ long and is discretised with a mesh of $720$ grid points in the streamwise direction and $240$ grid points in the two cross-flow directions. The immersed particle is initially spherical with and initial diameter $d$ equal to $d = D/5$ (blockage ratio $d/D=0.2$), which corresponds to $48$ Eulerian grid points per particle diameter ($\Delta x_i =1/48$). We also performed a simulation with a finer mesh of $72$ points per particle diameter ($\Delta x_i =1/72$) and found no appreciable difference, which ensures the grid independency of the results on the coarser grid resolution which is used for all the other cases. In all our simulations, the time-step is chosen to ensure a CFL number equal to $0.2$. The present simulations, with a resolution of $48$ points per diameter, require approximately $4$ to $16$ days with $400$ computational cores to reach the particle final equilibrium position, depending on the value of the elastic modulus $G$.

The problem at hand is defined by two non-dimensional parameters: the Reynolds and the Weber numbers. The former is defined as $Re = \rho D U_b/\mu^f$, where $U_b$ is the bulk velocity across the domain. In our simulations, $U_b$ is kept constant,ensuring a constant mass flux, by applying a varying pressure gradient driving the flow through the domain. We vary the Reynolds number between $100$ and $400$ to study the inertial migration of the particle in a range of Reynolds number of interest for emerging inertial microfluidics applications \cite{amini2014}. The Weber number is defined as $We = \rho U^2_b / G$ and is varied between $0.125$ and $4$, ranging from an almost rigid particle ($We = 0.125$) to an highly deformable one ($We = 4$). Two additional non-dimensional parameters are the density and viscosity ratios which are fixed equal to $1$ in all our simulations. Finally, the particle is initially positioned at a distance $r \approx 0.3 R$ from the center of the pipe (where $r$ denotes the radial position of the particle center), except in two additional simulations where the particle is initially located at $r \approx 0.75R$ and $r \approx 0.05R$ to study the effect of the initial position on the results. 
The full set of simulations and parameters considered in the present study is summarized in table~\ref{tab:allcases}.

\section{Results} 
Here, we study the migration of an initially spherical hyperelastic particle in a Hagen--Poiseuille flow of a Newtonian fluid for different Reynolds and Weber numbers and for various solid to fluid viscosity ratios. Figure~\ref{fig:box} shows the particle migration for the case $Re= 400$ and $We = 1$; we observe that the particle deforms and undergoes a lateral displacement while traveling downstream through the pipe, finally focusing at the pipe centerline, in agreement with previous observations \cite{villone2016}. The same general behavior is found for all the other cases we have investigated, with the full migration process and the final particle equilibrium position being determined by the interplay between the different opposing forces acting on the particle.

\begin{figure}[t]
\centering
\includegraphics[width=0.99\linewidth]{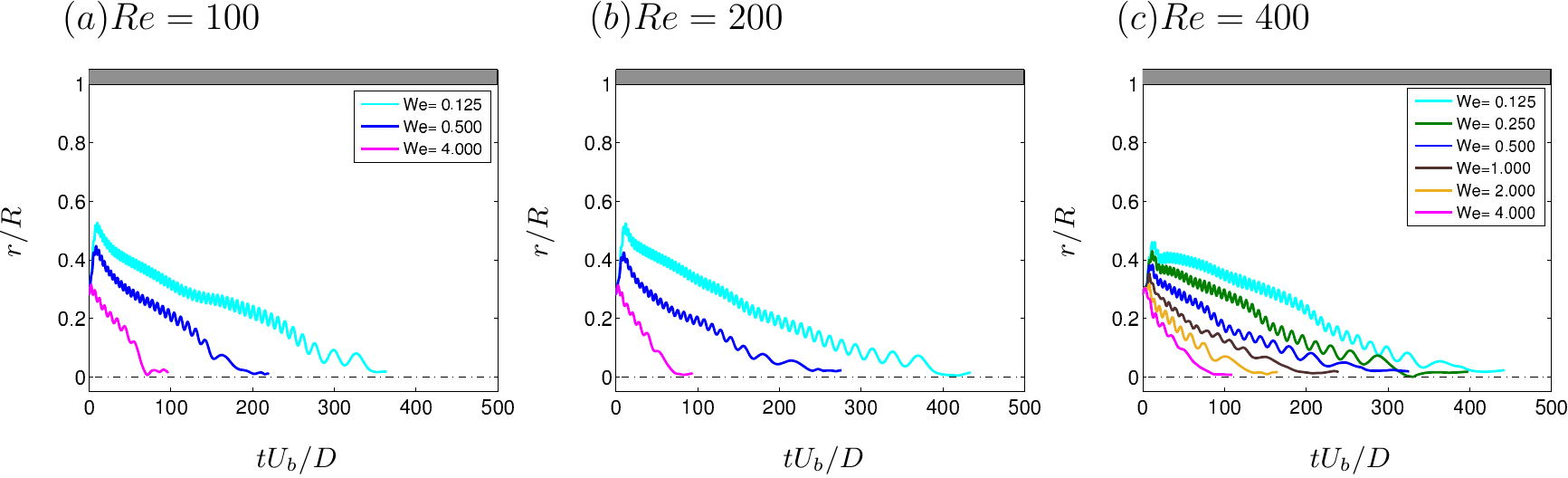}
\caption{(Colour online) The time history of the particle radial position $r$ at different Weber numbers $We$ for various Reynolds numbers $Re$, as indicated in the legend. The radial position $r$ is normalized by the radius of the pipe $R$ whereas time is scaled by $D/U_b$.}
\label{fig:Re_time} 
\end{figure}
\begin{figure}[t]
\centering
\includegraphics[width=0.99\linewidth]{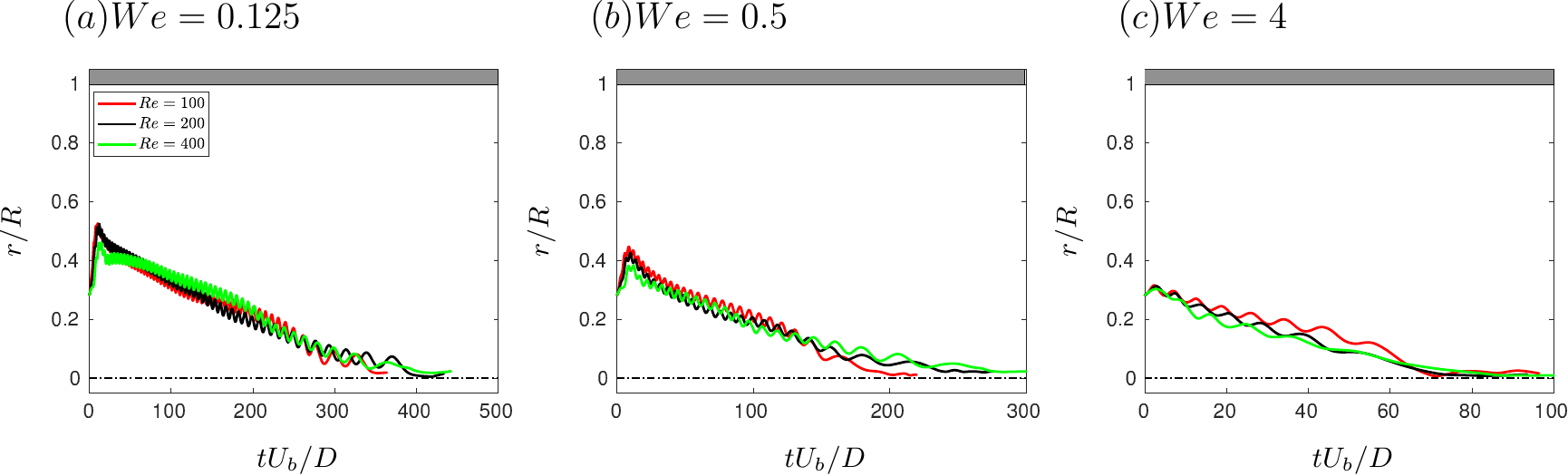}
\caption{(Colour online) The time history of the particle radial position $r$ at different Reynolds numbers $Re$ for various Weber numbers $We$, as indicated in the legend.}
\label{fig:We_time} 
\end{figure}
\begin{figure}[b]
\centering
\includegraphics[width=0.66\linewidth]{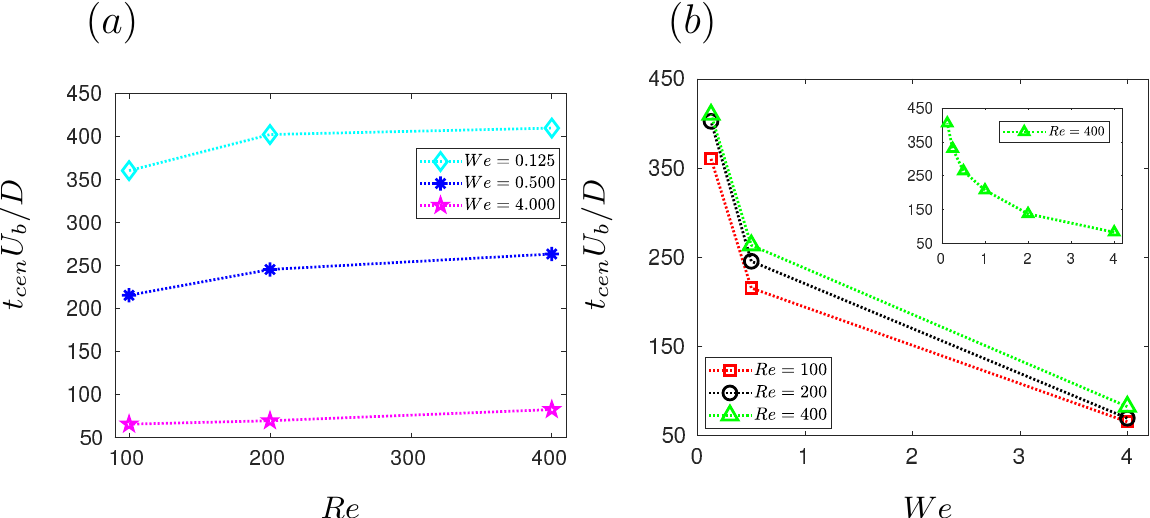}
\caption{(Colour online) Normalized time $t_{cen}$ needed to reach the final equilibrium position versus (a) the bulk Reynolds number $Re$ for 
different $ We$ and (b) the Weber number $We$ for different $Re$.}
\label{fig:t_eq} 
\end{figure}

The time history of the particle radial position, $r$, measured from the center of the pipe, is illustrated in figure \ref{fig:Re_time} for different Reynolds and Weber numbers. All the particles are released from the same initial position, $r=0.2828R$, and then migrate towards the center of the pipe, where they settle and reach their equilibrium position on the symmetry axis ($r \approx 0$). In the figure, we can also observe that the particle first displaces from its initial position towards the wall, and only after some time ($t U_b/D \geq 10$) starts migrating towards the pipe center. The first outer displacement is caused by the fact that the particle initial shape is spherical, and according to the famous Segr{\`e}-Silberberg effects \cite{segre1961}, rigid spherical particles tend to migrate towards the walls and settle down at a distance of approximately $r \approx 0.6R$. This equilibrium position originates from the balance of the force coming from the mean shear of the velocity profile pulling the particle towards the wall and the pressure pushing it towards the center. However, after some time the particle deforms, and starts displacing towards the pipe center, where it will settle due to the local zero shear rate of the mean velocity profile. The time needed to the particle to invert its motion from the motion towards the wall to the one towards the center decreases with the Weber number, i.e. for high $We$ the particle deformation is fast and the spherical shape is lost within a short time. Finally, from figure \ref{fig:Re_time} we can notice that for all the Reynolds number considered, as the Weber number increases, the whole migration process is faster and the final equilibrium position is reached in a shorter time than for the rigid particles cases with low $We$.

Next, we investigate the effect of the Reynolds number on the dynamics and equilibrium position of the particle at fixed Weber numbers. We show the time history of the particle motion from its initial radial location ($r/R=0.2828$) to the final equilibrium position ($r/R \simeq$ 0) for different Reynolds number in figure \ref{fig:We_time}(a-c) at three different Weber numbers, $We=0.125$, $0.5$ and $4$. Our results clearly show that both the migration dynamics and the final equilibrium position of a deformable particle is almost independent of the Reynolds number, at least in the range considered here (i.e., $ 100 \leq Re \leq 400$), thus suggesting that the dynamics of deformable particles is mainly controlled by the value of the particle elastic modulus $G$.

To further understand the effect of the Reynolds $Re$ and Weber $We$ numbers on the dynamics of the deformable particle, we measure the time needed for the particle to reach its final focal position, $t_{cen}$, for all the cases studied. Figure \ref{fig:t_eq} shows the normalized time to equilibrium $t_{cen}$ as a function of the Reynolds number $Re$ (panel a) and of the Weber number $We$ (panel b). The figure confirms our previous observations that the equilibrium time is strongly affected by $We$ and weakly by $Re$. In particular, $t_{cen}$ slightly increases with the Reynolds number $Re$ while substantially decreases with the Weber number $We$ (for example at $Re=200$, $t_{cen}$ reduces by almost $8$ times when increasing $We$ from $0.125$ to $4$). Finally, in the inset of figure \ref{fig:t_eq}(b) we report all our data for the case at $Re = 400$ which show that the decay of $t_{cen}$ with $We$ is smooth and monotonic.

\begin{figure}[t]
\centering
\includegraphics[width=0.33\linewidth]{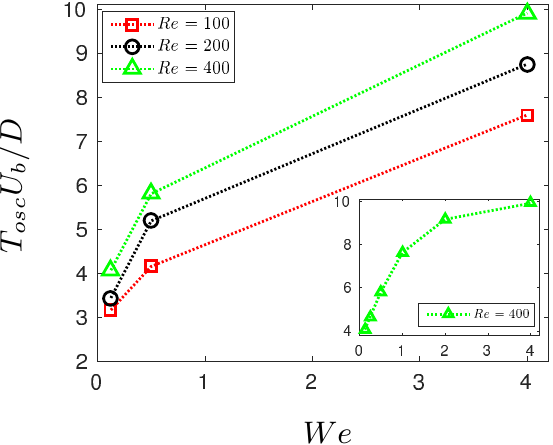}
\caption{(Colour online) Normalized period of the particle oscillatory motion, $T_{osc} U_b/D$, as a function of the Weber number $We$ for different $Re$.}
\label{fig:period_T} 
\end{figure}
In both figures \ref{fig:Re_time} and \ref{fig:We_time} we can observe that the particle migration towards the center presents oscillations. Also, the period of the particle oscillations is approximately the same for the different $Re$ at constant $We$ (figure \ref{fig:We_time}), while the period increases when increasing the Weber number (figure \ref{fig:Re_time}), thus suggesting that the oscillations are due to elastic effects only. Figure \ref{fig:period_T} reports the normalized period of the oscillation, $T_{osc} U_b/D$,  as a function of the Weber number $We$; $T_{osc}$ is the mean period of the oscillation averaged over the first half of the particle migration process, where the period remains approximately constant, while the second half is neglected since $T_{osc}$ slightly changes and increases when the particle approaches the final equilibrium position. As shown in figure \ref{fig:period_T}, the period $T_{osc}$ monotonically and non-linearly increases with $We$; in particular, the growth of $T_{osc}$ is very fast for low $We$, eventually almost saturating at higher $We$. Also, from the figure we observe that $T_{osc}$ slightly increases with the Reynolds number, altough the growth is small compared to the one due to the Weber number $We$.

\begin{figure}[t!]
\centering
\includegraphics[width=0.99\linewidth]{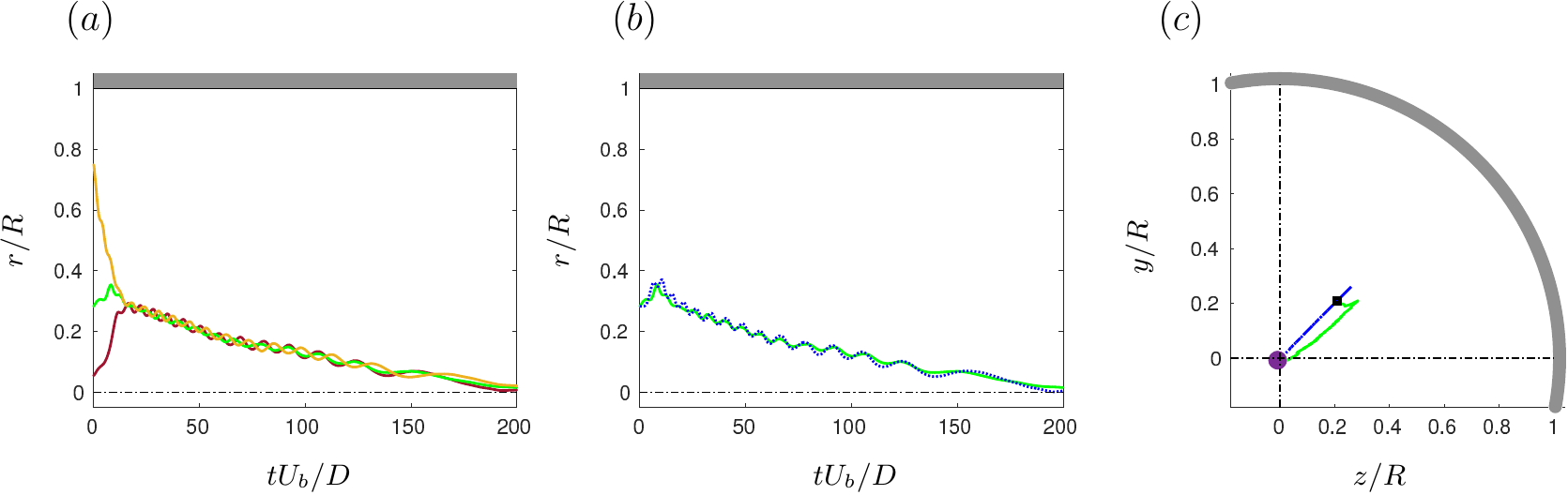}
\caption{(Colour online) The time history of the radial position of the deformable particle released at (a) three different initial positions and (b) with two different initial velocities at the same Reynolds and Weber numbers: $Re= 400$ and  $We= 1$. (c) Trajectory of the particle in the pipe cross-section ($y-z$ plane) for the cases reported in panel (b).}
\label{fig:Re400}
\end{figure}
The previous results are general and do not depend on the particle initial position and velocity. We show this by examining the effect of the radial initial position and velocity on the dynamics and final equilibrium position of the particle at a fixed Reynolds and Weber numbers. Figure \ref{fig:Re400}(a) shows the time evolution of the radial position of a particle released from three different radial positions ($r/R = 0.0544, 0.2828, 0.7487$) at $Re= 400$ and  $We= 1$. The figure clearly shows that, the particle always migrates towards the center of the pipe, independently of its initial radial position. Interestingly, after short initial transients during which the single particles reach the radial position $r \approx 0.29R$, the trajectories collapse into a single oscillating curve for $t U_b/D \gtrsim 15$. Figure \ref{fig:Re400}(b) reports the time evolution of the radial position of a particle released from $r/R = 0.2828$ with two different initial conditions: the solid green line shows the case with the particle released stationary in a fully developed pipe flow (a Poiseuille velocity profile), while the blue dashed line the case where both the particle and fluid are released at rest, i.e.,\ with zero velocity. The two particle trajectories are very similar, which indicates the independence of the results on the particle initial conditions. Although the initial particle velocity does not affect the trajectory of the particle radial position, it influences the motion in the cross-section of the pipe, as reported in figure \ref{fig:Re400}(c). Indeed, only the latter case exhibits a straight radial motion from the particle initial position towards the pipe center, while in the former case a non-zero azimuthal velocity is observed. As already mentioned, please note that we have also verified that this behaviour is independent of the grid resolution used.

\begin{figure}[t!]
\centering
\includegraphics[width=0.99\linewidth]{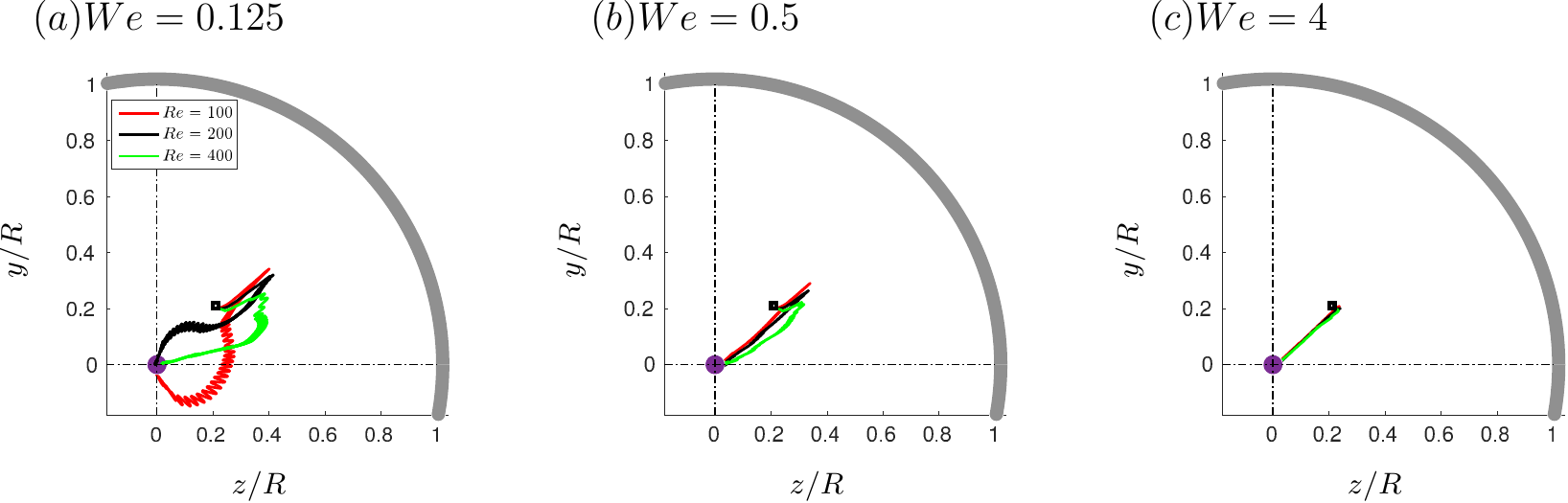}
\caption{(Colour online) Trajectory of the particle in the pipe cross-section ($y-z$ plane) at different Reynolds numbers $Re$ and for various Weber numbers $We$. The black squares and purple circles show the initial and the final equilibrium positions, respectively.}
\label{fig:We_pos} 
\end{figure}
The trajectories of the particles in the $y-z$ plane from their initial position till the final equilibrium are depicted in figure \ref{fig:We_pos} for all the Reynolds, $Re$, and Weber, $We$, numbers considered. The initial and final equilibrium positions of the particles are marked with black squares and purple circles in the figure. Clearly, the results show that independently of the Reynolds numbers $Re$ under investigation, the particle path length decreases with the Weber number $We$. Indeed, for the highest $We$ considered (figure \ref{fig:We_pos}(c)), the particle moves straight inwards towards the center of the pipe along the radial direction, with all the trajectories for different Reynolds number $Re$ collapsing into a single line. On the other hand, as the Weber number decreases and the particle rigidity increases, the motion becomes more complex, with the particle initially moving towards the pipe walls, and then shooting back towards the center of the pipe. Moreover as the Weber number decreases, we can observe that particles at different Reynolds number exhibit different cross-sectional trajectories, as clearly shown in figure \ref{fig:We_pos}(a).

\begin{figure}[t!]
\centering
\includegraphics[width=0.7\linewidth]{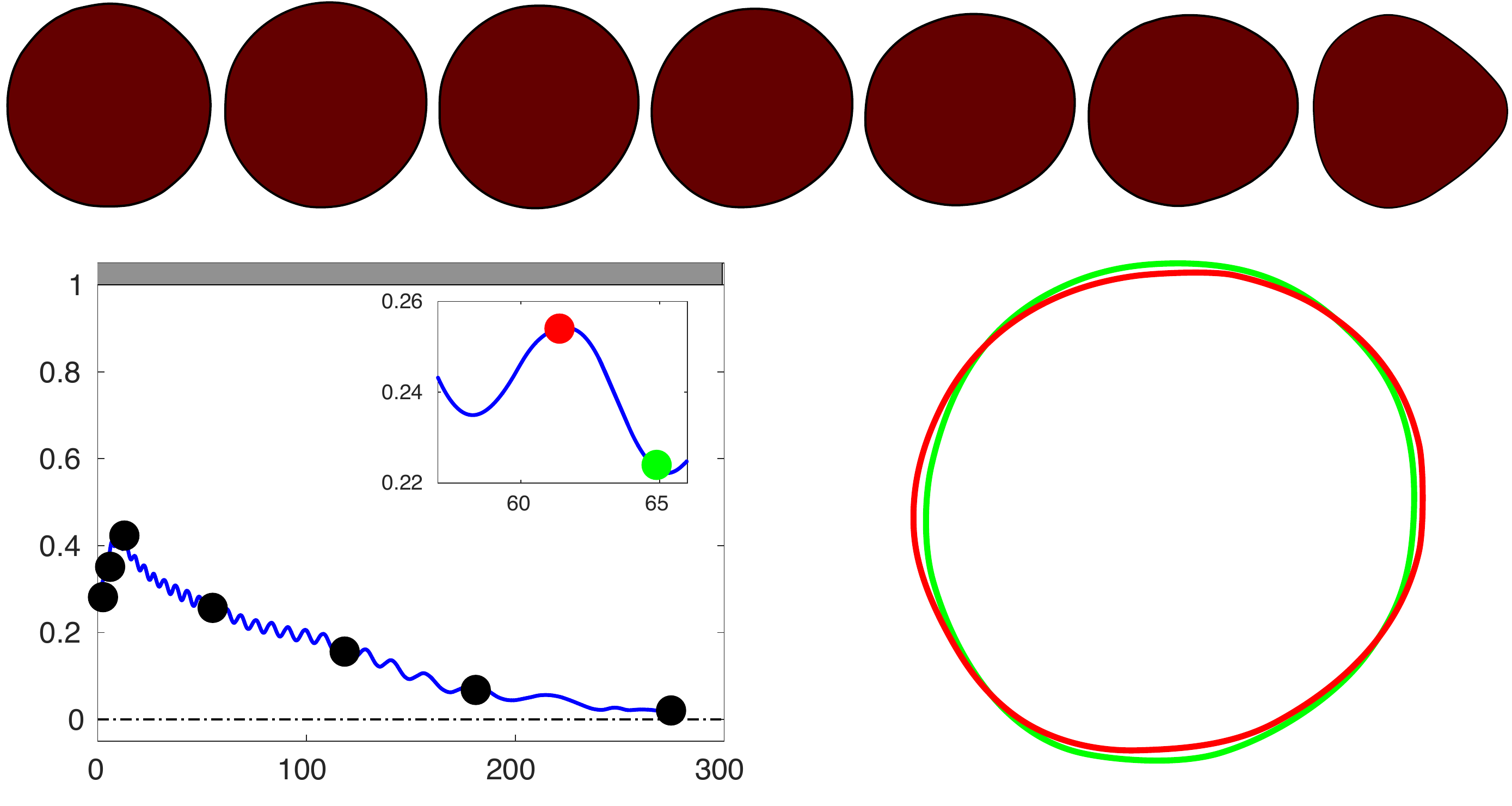}
\put(-365,63){{\rotatebox{90}{$r/R$}}}
\put(-270,-10){{$t U_b/D$}}
\caption{(Colour online) Cross section of the deformed shape of an hyper-elastic particle at $Re=200$ and $We=0.5$ at seven different time instants, marked in the time history of the radial position with black circles, from the initial to the equilibrium position. The red and green lines represent the change of particle shape during an oscillation.}
\label{fig:particle_time}
\end{figure}

During the migration, the particle deforms and changes shape due to the different hydrodynamic stresses imposed by the flow at the different radial positions, as depicted in figure \ref{fig:particle_time}. In particular, we observe that the particle, initially spherical, deforms into an asymmetric shape during its migration motion and finally assumes an axial-symmetric configuration when reaching the center of the channel. Indeed, the particle asymmetric shape and the related non-uniform mean velocity profile are crucial for the particle migration process \cite{kaoui2008}. This is detailed in figure \ref{fig:particle_time}, where we display the shapes assumed by the particle during their motion for the case with $Re=200$ and $We=0.5$; the leftmost panel represents the particle initial shape (at time $t=0$), the rightmost panel the particle final shape at the equilibrium (at time $t=t_{cen}$), while the middle panels the shape assumed during the migration. In the first part of the process ($t \lesssim 5 D/U_b$), the shape is approximately spherical, and the particle migrates towards the wall; when the deformation builds up and the particle becomes asymmetric, the particle reverts its motion and displaces towards the pipe centerline. As already stated above, this transient shape is non-symmetric due to the different shear rates at various wall-normal distances and the finite size of the particle. On the other hand, the particle assumes an axial-symmetric bullet-like shape when it reaches the pipe center due to the symmetry of the velocity profile. Similar shapes were observed both numerically and experimentally by various authors in the past for deformable vesicles, particles, capsules and cells \cite{vitkova2004, kaoui2008, villone2016, raffiee2017, schaaf2017, mokbel2017}, in the limit of vanishing Reynolds numbers, as this corresponds to a minimum of the elastic energy. Figure \ref{fig:particle_time} also shows how the particle shape changes during its oscillatory motion. In particular, the figure reports the particle shape over half a period of osicllation, i.e.\ $T_{osc}/2$. As expected, we observe that the particle shape slightly changes during the oscillation process; however, the amplitude of the oscillation is small, around $1\%$ of $R$.

\begin{figure}
\centering
\includegraphics[width=0.5\linewidth]{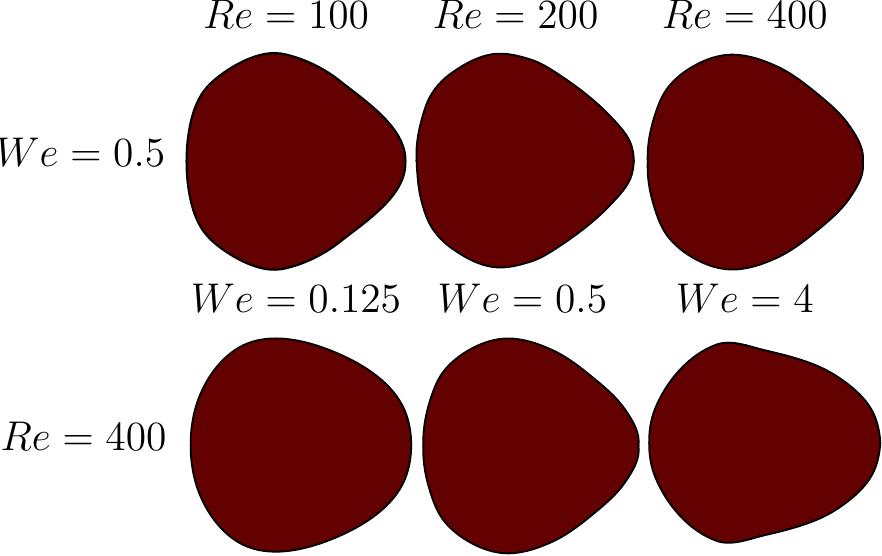}
\caption{(Colour online) Cross section of the deformed shape of an hyper-elastic particle for (top) three different Reynolds numbers and at the same Weber number  $We=0.5$ and for (bottom) three different Weber numbers and at the same Reynolds number  $Re=400$ at the final equilibrium state.}
\label{fig:particle} 
\end{figure}
In figure \ref{fig:particle} we show the effect of the Reynolds and Weber numbers on the equilibrium shape of the particle. From the figures we note that, the equilibrium shape of the particle is only slightly affected by variations of the Reynolds number $Re$ (top row in figure \ref{fig:particle}), while the shape changes with the Weber number (bottom row in figure \ref{fig:particle}); in particular, at high Weber numbers $We$, the particle becomes more elongated than for low $We$. These results are consistent with what previously observed in figures \ref{fig:Re_time} and \ref{fig:We_time}.

\begin{figure}[t]
\centering
\includegraphics[width=0.99\linewidth]{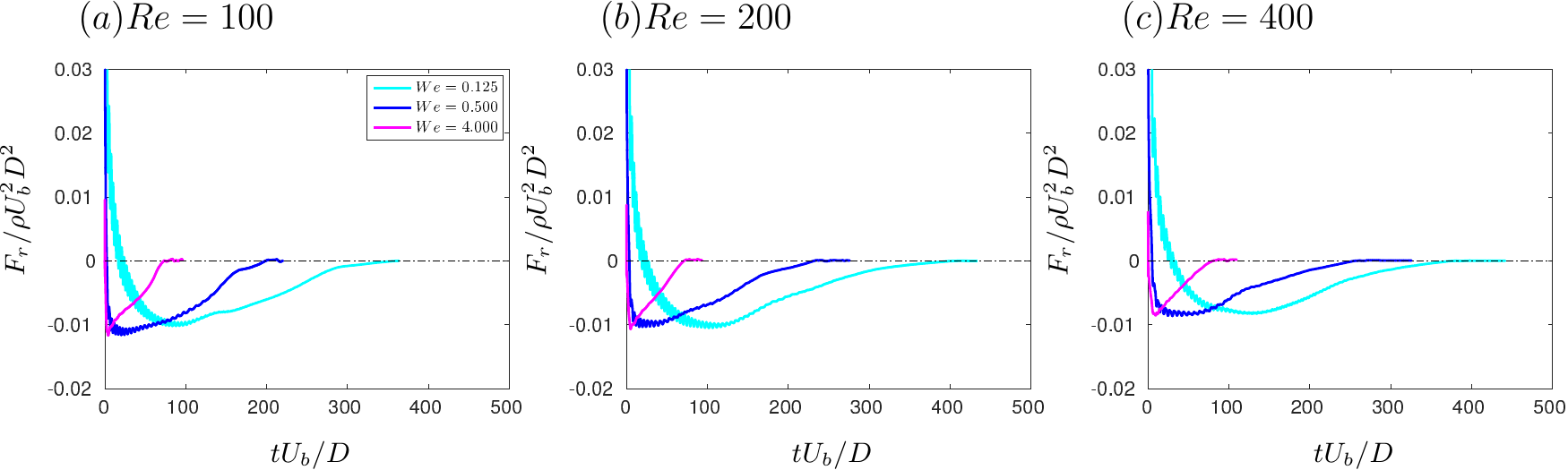}
\caption{(Colour online) The time history of the normalized total radial forces $F_r$ acting on the particle at different Reynolds numbers $Re$ and for various Weber numbers $We$.}
\label{fig:Fn}
\end{figure} 
Next, we discuss the forces acting on the particle to explain its migration dynamics. In the case of a neutrally buoyant rigid particle suspended in a Newtonian fluid, the particle migration and final equilibrium position are determined by two opposing forces acting on the particle: these forces are the wall induced lift force that pushes the particle away from the wall and the shear-gradient induced lift force that drives the particle towards the wall, the latter resulting from the rigid particle resistance to deformation \cite{martel2014, stoecklein2018}. When the particle is deformable, its dynamics is further complicated by the additional force originating from the particle shape deformation, which depends on the elastic properties of the material (e.g., the elastic modulus $G$) \cite{raffiee2017, hadikhani2018}. Note that, the particle deformation also affects and modifies the other two forces.

\begin{figure}[t]
\centering
\includegraphics[width=0.33\linewidth]{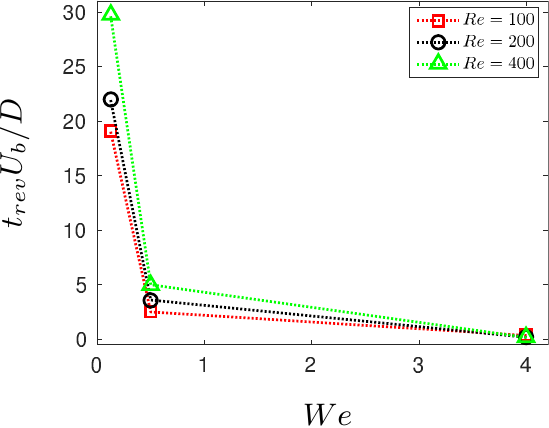}
\caption{(Colour online) Normalized time needed to reverse the sign of the total force acting on the particle, $t_{rev}$, versus the Weber number $We$ for different $Re$.}
\label{fig:t_rev} 
\end{figure}

Figure \ref{fig:Fn}(a-c) shows the total force acting on the particle for different Reynolds numbers and for various Weber numbers, as a function of time. We observe that, the total force is positive at the beginning of the simulation, then changes sign and finally vanishes to zero at later times. This is consistent with what observed in figure \ref{fig:Re_time} which shows that the particle first moves towards the walls and then changes direction to move towards the center. This behaviour is due to the fact that the particle is initially spherically and thus tends to focus around $0.6R$, but then starts deforming and the asymmetric shape drives the particle to the centerline; this is not an instantaneous process due to the finite inertia of the flow and the elastic timescale. In particular, the migration timescale mostly changes with the Weber number, as shown in the figure: for small $We$ (low deformability) the process is slow, while for high $We$ (high deformabilty) the process is fast. We conjecture that, for $We=0$ (rigid particle) the process needs an infinite time and thus would never happen, leading to the particle focusing at $0.6R$; indeed, we can observe in figure \ref{fig:t_eq}(b) that the time needed to focus at the centerline strongly increases when reducing the Weber number, apparently diverging for $We=0$. This is also shown in figure \ref{fig:t_rev} where the time $t_{rev}$ when the total force is first null, $F_r=0$, i.e.\ it changes sign, is reported as a function of the Weber number $We$ and for different Reynolds numbers $Re$. As observed for the focusing time in figure \ref{fig:t_eq}(b), $t_{rev}$ decreases with the Weber number and only slightly increases with the Reynolds number. The growth of $t_{rev}$ with the Reynolds number is due to the increase of the particle inertia, which requires longer times to invert its motion. Also, $t_{rev}$ tends to zero for $We \rightarrow \infty$ and tends to infinity for $We \rightarrow 0$.

\begin{figure}[t]
\centering
\includegraphics[width=0.99\linewidth]{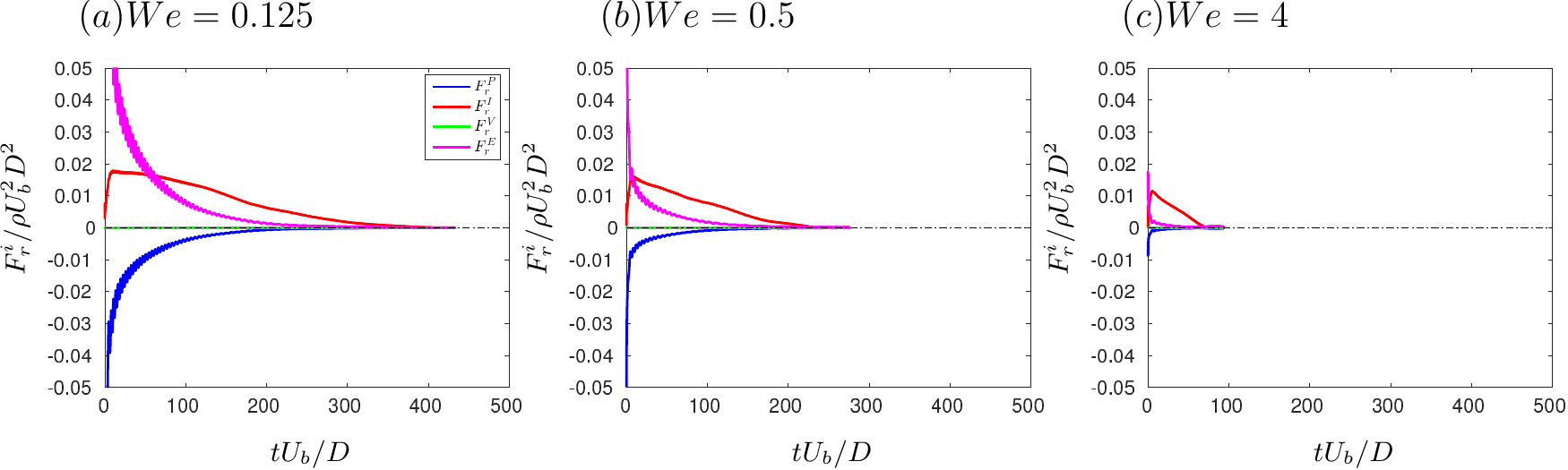}
\caption{(Colour online) The time history of the normalized radial force components $F_r^i$ acting on the particle for different Weber numbers and a fixed Reynolds number $Re=200$.} 
\label{fig:Fri}
\end{figure} 
Finally, we compute the average over the particle volume of each term appearing in the momentum conservation equation , i.e. the pressure, non-linear inertial, viscous and elastic contributions to equation (\ref{eq:NS}c); in particular, the force components in the $i$-th direction are computed as
\begin{equation}
\label{eq:force}
F_i^P = -\int_{\mathcal{V}^s} \frac{\partial P}{\partial x_i}~d\mathcal{V}, \;\;\;\; F_i^I = -\int_{\mathcal{V}^s} \frac{\partial u_i^s u_j^s}{\partial x_j}~d\mathcal{V}, \;\;\;\; F_i^V = \int_{\mathcal{V}^s} \mu^s \frac{\partial}{\partial x_j} \left( \frac{\partial u_i^s}{\partial x_j} + \frac{\partial u_j^s}{\partial x_i} \right) ~d\mathcal{V}, \;\;\;\; F_i^E = \int_{\mathcal{V}^s} G \frac{\partial \xi_{ij}}{\partial x_j}~d\mathcal{V}.
\end{equation}
Note that, the sum with sign of these terms gives the rate of change of the particle velocity; also, by applying the Gauss's theorem we can rewrite $F_i^P$ and $F_i^V$ as surface integrals, as in the classical definitions of pressure and viscous contributions to the lift and drag forces on an object. For a completely solid particle, the force balance is given by the wall-induced lift force ($F_i^P$) arising from the interaction of the particle and the neighboring wall pushing towards the channel centerline, and a shear gradient lift force pulling the particle towards the wall. The latter arises from the fact that the particle, being rigid, cannot sustain the torque generated by the mean shear gradient ($F_i^V$ and $F_i^I$) and thus starts to rotate. In the case of a deformable particle, one additional force is present ($F_i^E$): the elastic force inside the particle. Figure \ref{fig:Fri} shows these different terms for the case with $Re=200$ and different $We$; first, we observe that the viscous force is approximately zero and does not contribute to the momentum balance, which is due to the fact that the particle can actually deform. From the figure, we can also infer that the pressure term is negative, similarly to the rigid particle case, thus pushing the particle towards the centerline, while the elastic and non-linear terms are positive, thus pushing the particle towards the wall. In the initial part of the particle dynamics, the non-linear term is zero and slowly grows, while the two dominant force contributions are the pressure and elastic forces, the latter being the largest, so that, overall, these induce the movement of the particle towards the wall. As the particle deforms and aligns with the mean shear, the elastic force decreases rapidly and the non-linear term becomes the dominant one. However, deformation also prevents particle rotation, thus significantly reducing the inertial force towards the wall. As a consequence, there remains only a pressure-induced force towards the center. As the deformation is faster for larger values of the Weber number, the lateral migration velocity increases with $We$, i.e.\ the focusing time decreases with deformability. From figure \ref{fig:Fri} we observe that, not only the total force is null when the particle reaches the centerline, but also all the force components vanish due to the symmetry; this is very different from the Segr{\`e}-Silberberg equilibrium position where the total force is zero, but the single force components are not. This discussion holds for all the Weber numbers considered in this study; however, as $We$ increases the time needed to reach the equilibrium strongly reduces, consistently with the fact that for $We=0$ this time should go to infinity.

\begin{figure}
\centering
\includegraphics[width=0.33\linewidth]{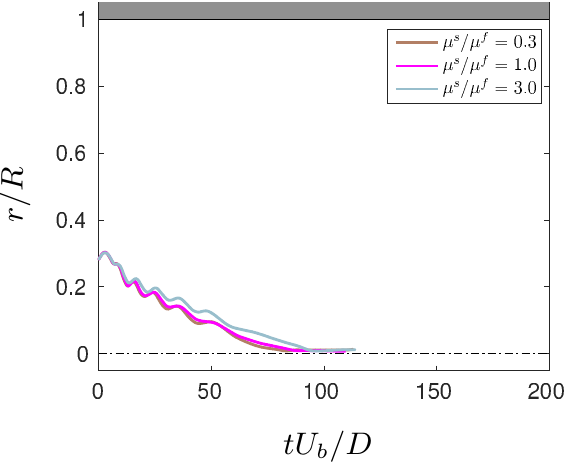}
\caption{(Colour online) Time history of the radial position of the deformable particle for three different solid/fluid viscosity ratio at the same Reynolds and Weber numbers: $Re= 400$ and  $We= 4$.}
\label{fig:Re400_visc}
\end{figure}
Finally, in this last section, we briefly assess the effect of the ratio of the solid viscosity $\mu^s$ to the fluid viscosity $\mu^f$. Here, we focus our analysis on a single Reynolds number $Re=400$ and Weber number $We=4$, and we consider three different values of the solid viscosity ratio, covering one order of magnitude: $\mu^s= 0.3$, $1$ and $3 \mu^f$. Figure \ref{fig:Re400_visc} reports the time history of the particle radial position; in general, the results suggest that high solid viscosity makes the particle effectively more rigid, thus exhibiting a behavior more similar to that of stiffer particles, i.e.\ the time needed to reach the equilibrium position in the channel centerline increases with $\mu^s$; conversely, low values of $\mu^s$ increase the particles deformation, thus reducing the time $t_{cen}$ needed to reach the final equilibrium position. Note that, the particle trajectory is only weakly modified by this parameter (compared to the modification induced by $We$), at least for the range of parameters considered in this work. These results are in agreement with what previously observed by Rosti and Brandt in Ref.\ \cite{rosti_brandt_2018a}.

\section{Conclusions and final remarks} 
We study the dynamics of an hyperelastic neo-Hookean deformable particle suspended in a Newtonian fluid in a straight pipe with circular cross-section. Different finite Reynolds and Weber numbers are considered, in order to evaluate the effects of inertia and elasticity on the particle focusing. The problem is studied numerically through an extensively validated fully Eulerian formulation based on the one-fluid formulation where a single set of equations is solved for both the fluid and solid phases.

We find that the particle deforms and undergoes a lateral displacement while traveling downstream through the pipe, always focusing at the pipe centerline. While the particle final equilibrium position is independent of the Reynolds and Weber numbers considered, its migration dynamics strongly depends on the particle elasticity while it is only slightly affected by the Reynolds number. In particular, the migration is faster as the elasticity increases, with the particle reaching the final equilibrium position at the centerline in shorter times when more deformable.

When the particle is injected in the flow at some intermediate position, it first moves towards the walls aiming for the famous Segr{\`e}-Silberberg equilibrium position of rigid particles ($0.6R$). However, as soon as the particle starts deforming, it changes direction of motion and starts migrating towards the centerline. Indeed, the particle, initially spherical, deforms into an asymmetric shape due the non-uniform shear, which ultimately causes its movement to the centerline. Indeed, when the final equilibrium position at the pipe center is reached, the particle assumes an axisymmetric bullet-like shape that enforces the equilibrium.

In order to explain the migration dynamics, we analyze the force acting on the particle and found that the total force is first positive, thus pushing the particle towards the wall, then becomes negative, causing the particle migration to the centerline, and finally vanishes to zero when the particle reaches the final equilibrium position. We decompose 
the total force acting on the particle in different contributions, i.e., the viscous, pressure, inertial and elastic contributions. We found that the viscous force is negligible, the pressure term is responsible for pushing the particle towards the centerline, similarly to the rigid particle case, while the elastic and inertial ones are opposing the movement trying to pull the particle towards the wall. However, this opposition is only limited to the initial transient phase, as particle deformation aligns it with the local shear, reducing the elastic force, and, more importantly, prevents particle rotation, which quenches the inertial force towards the wall. Indeed, the migration is faster, when the deformation is faster, i.e.\ for larger values of the Weber number.

Finally, we show the effect of the solid to fluid viscosity ratio and show that high solid viscosity makes the particle effectively more rigid, so that it requires a longer time to reach the equilibrium position when compared to cases with low values of solid viscosity. Also, we observe that the effect of the solid viscosity is smaller than that of the Weber number.

\section*{Acknowledgments}
The authors were supported by the European Research Council Grant no. ERC-2013-CoG-616186, TRITOS and by the Swedish Research Council Grant no. VR 2014-5001. The authors acknowledge computer time provided by the Swedish National Infrastructure for Computing (SNIC). Dhiya Alghalibi would like to gratefully acknowledge his graduate scholarship from the Iraqi Ministry of Higher Education and Scientific Research via University of Kufa.

\section*{Appendix: the rigid particle limit}
\begin{figure}[t]
  \centering
  \includegraphics[width=0.33\linewidth]{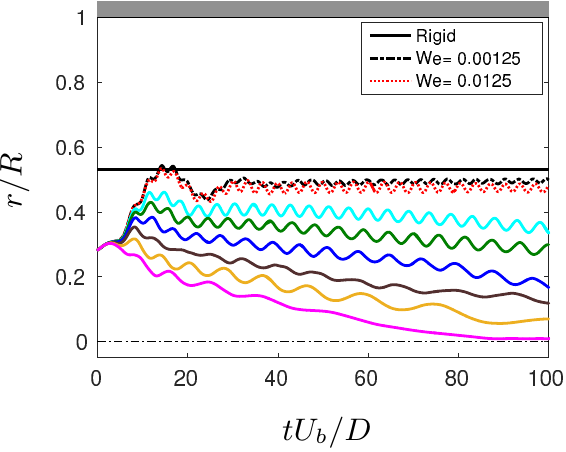}
  \caption{The time history of the particle radial position $r$ at different Weber numbers $We$ at constant Reynolds numbers $Re=400$, as indicated in the legend. The radial position $r$ is normalized by the radius of the pipe $R$ whereas time is scaled by $D/U_b$.}
  \label{fig:appendix} 
\end{figure}

All the cases studied in the rest of the manuscript shows that the final equilibrium position of a deformable particle is the centerline and that the time needed to go towards the center grows as the particle becomes more rigid. Here, we show some additional results for very rigid particles (high modulus of transverse elasticity $G$) to enforce this concept and to show that the limit of fully rigid particle can be approached and the Segr{\`e}-Silberberg equilibrium position recovered. Figure \ref{fig:appendix} displays the results for a particle at $Re=400$ and all the $We$ considered in the study plus two additional cases with very rigid particles (yet not fully rigid). As shown in the figure, the equilibrium position approaches the value predicted for a fully rigid particle as $G$ increases, tending to the value $0.53R$ predicted by Asmolov \cite{asmolov1999} for a finite size rigid particle as the one considered here. Although in the chosen time frame the most rigid particle does not move towards the centerline yet, it  will eventually deform and will start moving towards it. This process takes place in a time which strongly grows with $G$, as also shown in the main text of this manuscript. Thus, we can consider a perfectly rigid particle as a deformable one with infinite modulus of transverse elasticity that requires an infinite time to deform and start displacing towards the channel centerline.

%

\end{document}